\documentclass{article}
\usepackage[utf8]{inputenc}
\usepackage{authblk}
\usepackage{amsmath}
\usepackage{amssymb}
\usepackage{amsthm}

\usepackage{mathtools}
\usepackage{graphicx}
\usepackage{hyperref}
\usepackage{float}
\usepackage{anysize} 
\usepackage{subcaption}
\usepackage{xcolor}
\begin{document}

\title{Vectorial active matter on the lattice: polar condensates and nematic filaments}
\author[1,2,*]{Josu\'e Manik Nava-Sede\~no}
\author[3]{Haralampos Hatzikirou}
\author[1,4]{Anja Vo{\ss}-B\"ohme}
\author[1]{Lutz Brusch}
\author[1,*]{Andreas Deutsch}
\author[5,*]{Fernando Peruani}

\affil[1]{Technische Universit\"at Dresden, Center for Information Services and High Performance Computing, N\"othnitzer Stra{\ss}e 46, 01062 Dresden, Germany}
\affil[2]{Department of Mathematics, Faculty of Sciences, Universidad Nacional Aut\'onoma de M\'exico, Circuito Exterior, Ciudad Universitaria, 04510 Mexico City, Mexico}
\affil[3]{Mathematics Department, Khalifa University, P.O. Box 127788, Abu Dhabi, UAE}
\affil[4]{Fakult\"at Informatik/Mathematik, Hochschule f\"ur Technik und Wirtschaft, Dresden, Germany}
\affil[5]{Laboratoire de Physique Th\'eorique et Mod\'elisation, UMR 8089, CY Cergy Paris Universit\'e, 95302 Cergy-Pontoise, France}
\affil[*]{Corresponding authors. Email: manikns@ciencias.unam.mx}

\maketitle

\begin{abstract}
We introduce a novel lattice-gas cellular automaton (LGCA) for compressible vectorial active matter with polar and nematic velocity alignment. Interactions are, by construction, zero-range. 
For polar alignment, we show the system undergoes a 
phase transition that promotes aggregation with strong resemblance to the classic zero-range process (ZRP). 
We find that above a critical point, the states of a macroscopic fraction of the particles in the system coalesce into the same state, sharing the same position and momentum (polar condensate).    
For nematic alignment, the system also exhibits condensation, but there 
exist fundamental differences: a macroscopic fraction of the particles in the system 
collapses into a filament, where particles possess only two possible momenta. 
Furthermore, we derive hydrodynamic equations for the active LGCA model to understand the phase transitions and condensation that undergoes the system. We also show that generically the discrete lattice symmetries -- e.g. of a square or hexagonal lattice -- affect drastically the emergent large-scale properties of on-lattice active systems. 
The study puts in evidence that aligning active matter on the lattice displays 
new behavior, including phase transitions to states that share similarities to condensation models.
\end{abstract}


\section{Introduction}
Self-organized patterns of self-propelled entities are found at every scale in biology:  
from {\it in vitro} cytoskeletal extracts~\cite{schaller_polar_2010, sanchez_spontaneous_2012, nagai_collective_2015}, bacterial systems~\cite{dombrowski_self-concentration_2004, sokolov_concentration_2007, peruani_collective_2012,ariel_swarming_2015,beer_statistical_2019}, and insect swarms~\cite{attanasi_finite_2014} to schools of fish~\cite{ward_quorum_2008}, herds of mammals~\cite{ginelli_intermittent_2015}, 
and human crowds~\cite{moussaid_how_2011}. 
Collective phenomena are also observed  in artificial active systems such as synthetic swimmers~\cite{palacci_living_2013,theurkauff_dynamic_2012}, phoretic colloids~\cite{buttinoni_dynamical_2013, golestanian_designing_2007}, 
and active rollers~\cite{bricard_emergence_2013,bricard_emergent_2015,Kaisere1601469}. 
Active systems in the absence of torques -- often referred as scalar active matter -- can phase separate~\cite{barberis2019, digregorio_full_2018, fily_athermal_2012}.  
Other complex, self-organized,  collective motion patterns exhibit either polar or nematic (orientational)  order. Given the vectorial nature of orientational order, active systems sharing such a feature are often referred to as ``vectorial active matter''~\cite{gompper20202020}. 
It is widely believed that self-organized pattern formation in vectorial active matter requires  particles to interact via some type of velocity alignment mechanism~\cite{vicsek_collective_2012,marchetti_hydrodynamics_2013}, 
though alternative mechanisms exist~\cite{barberis2016large}, such as combined attraction-repulsion and active forces~\cite{caprini2023flocking}.   
The best known example of an active system with polar velocity alignment is the Vicsek model~\cite{vicsek_novel_1995}, where complex, active polar flows spontaneously emerge. 
A different  class of active systems is  obtained by changing the velocity alignment in the Vicsek model from polar to nematic~\cite{peruani_mean_2008,ginelli_large_2010}. 
Self-propelled (point-mass) particles with nematic alignment do not generate polar flows, but self-organize into high-density, nematic bands~\cite{ginelli_large_2010}. 
The theoretical understanding of vectorial active matter has been strongly based on the study of off-lattice Vicsek-like models by means of numerically costly, particle-based simulations~\cite{vicsek_novel_1995,ginelli_large_2010, gregoire2004onset,chate_simple_2006,chate_modeling_2008,kursten2020dry}  
and by the derivation of hydrodynamic equations for these systems~\cite{toner_long_1995,bertin_boltzmann_2006,ihle_kinetic_2011, kursten2021quantitative,peshkov_nonlinear_2012,peshkov_boltzmann_2014, grossmann_self_2013}; for comprehensive reviews, see~\cite{vicsek_collective_2012,marchetti_hydrodynamics_2013}.  
There is, however, a renewed interest in understanding the behavior of active matter on lattices given the suitability of efficient numerical schemes, and in particular due to the large repertoire of available analytical tools  to 
study lattice models that are expected to help to improve our theoretical understanding of active matter physics~\cite{peruani2011traffic, partridge2019critical, kourbane2018exact, manacorda2017lattice, solon2015flocking, nesbitt2021uncovering}. 

\begin{figure}
\centering
\includegraphics[width=\linewidth]{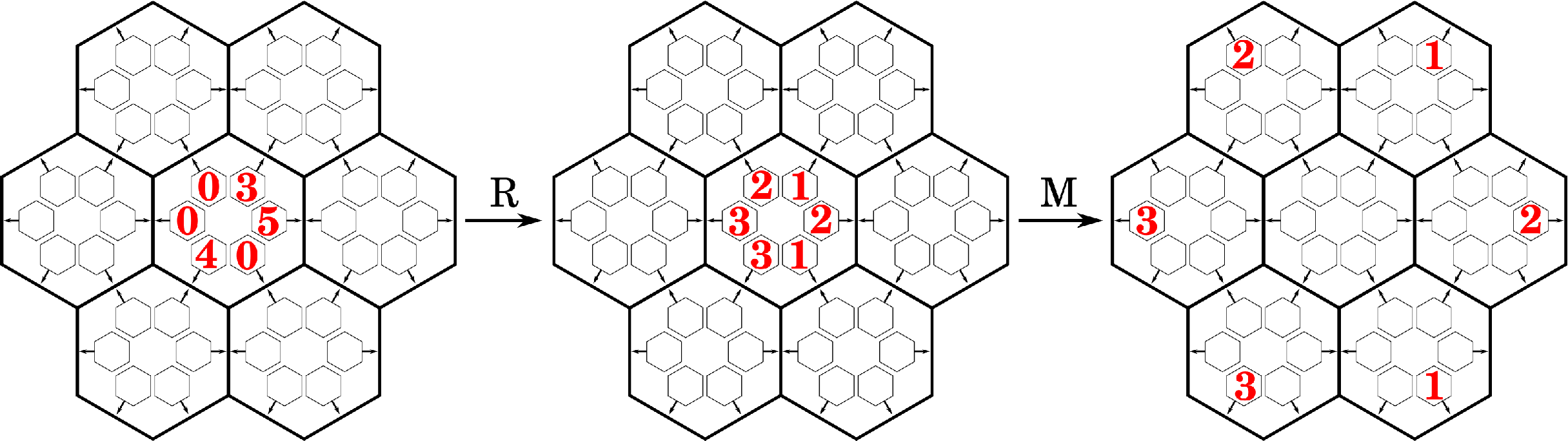}
\caption{LGCA dynamics. Particles reside in velocity channels within lattice sites. During the reorientation step (R), particles within each lattice site are  assigned new velocity channels according to the transition probability  $T_\vartheta\left(\mathbf{s}\right)$. 
Subsequently, during the migration step (M), particles are deterministically moved to neighboring lattice sites.}
\label{fig:dynamics}
\end{figure}

Here, we introduce an active lattice-gas cellular automaton (LGCA) for compressible (i.e. without volume exclusion)  vectorial active matter, including both polar and nematic velocity alignment, where 
interactions are, by construction, zero-range.  
We show that the introduced LGCA with polar alignment undergoes a non-equilibrium phase transition that is fundamentally different from the non-local, incompressible active LGCA, where the onset of order is characterized by the presence of travelling polar bands~\cite{bussemaker1997mean} as occurs in the  off-lattice Vicsek model~\cite{gregoire_onset_2004}.   
In contrast, in the present model (with polar alignment), the emergence of order occurs via a process by which a macroscopic fraction of the particles 
coalesce into the same state, and thus sharing the same position and momentum, i.e. a polar condensate. 
For nematic aligment, we find that the system displays a condensation, but 
with a fundamental difference: a macroscopic fraction of the particles in the system collapses into a nematic filament.
Note that recently, condensation behavior was found in scalar active matter in the presence of an external potential by a mechanism known edge-diffusivity~\cite{golestanian2019bose, meng2021magnetic}.  
In scalar active matter -- i.e. in self-propelled disks -- it has been observed that phase separation into dense and dilute phases can result from volume exclusion interactions among the particles, a phenomenon known as motility-induced phase separation (MIPS); see~\cite{cates2015motility} for a review).
Our study shows that in vectorial active matter, condensation behavior is based on velocity alignment and does not require an external potential or external field, even in the absence of volume exclusion interactions. 
Thus, the emergence of polar and nematic order in the active LGCA is fundamentally different from the onset of order in other models~\cite{vicsek_novel_1995, toner_long_1995, gregoire2004onset, partridge2019critical, nesbitt2021uncovering, kourbane2018exact, manacorda2017lattice, solon2015flocking, peruani2011traffic}. Similarly, phase separation in the active LGCA, that takes the form of a condensation, is fundamentally different from MIPS~\cite{cates2015motility}.
 At high interaction strengths and low effective temperatures, these models show a great variety of collective behaviors such as homogeneous, polar-ordered states, formation of spatially-extended clusters, emergence of long-range order, filament-linked clusters, phase separation into extended jammed clusters, the formation of moving bands, and rippling; whereas our active LGCA models consistently show only two phases: a disordered, homogeneous state, and an aligned, highly condensed state in either single lattice sites, or in filaments of a single lattice site in width, depending on the nature of the alignment interaction.
Furthermore, the hydrodynamic equations for the introduced 
LGCA 
allows us to show that generically the discrete lattice symmetries -- at least for square and hexagonal lattices -- have a strong impact on the emergent large-scale properties of on-lattice active systems. 
In summary, our study unveils the existence of condensed phases in vectorial active matter on the lattice, without attractive or excluded volume interactions.

\section{Model definition} 
We study an hexagonal LGCA, where each node possesses six velocity channels, $$\mathbf{c}_\vartheta=\left(\cos[(\vartheta-1) \frac{\pi}{3}]\,, \sin[(\vartheta-1) \frac{\pi}{3}]\right)$$ with $\vartheta \in [1,2,\cdots,6]$, which can be occupied by a number $s_{\vartheta}$ of cells moving in the direction specified by $\mathbf{c}_\vartheta$. 
In contrast to standard passive~\cite{frisch1986lattice} and active~\cite{bussemaker1997mean, deutsch2005cellular} LGCA that consider that each velocity channel of a node can either be unoccupied or occupied by at most one cell, i.e. $s_\vartheta\in\{0,1\}$ -- similar to the Pauli exclusion principle -- 
 here we assume particles without volume exclusion by letting any velocity channel $\mathbf{c}_\vartheta$ of a node contain an unrestricted number of particles $s_\vartheta\in\mathbb{N}$: i.e. many particle can share the same state (position and momentum). 
At each discrete time step, as illustrated in Fig.~\ref{fig:dynamics}, two processes take place: 

(i) Particles modify their velocity  by transitioning to a channel $\vartheta$, according to the transition probability $T_{\vartheta}$: 
\begin{equation}
\label{eq:reorient}
T_{\vartheta}=\frac{1}{Z}\exp\left[\beta\, H_{\vartheta}(\mathbf{r},k) \right]
\end{equation}
with $H_{\vartheta}(\mathbf{r},k)$ defined as: 
\begin{equation}
\label{eq:H}
H_{\vartheta}(\mathbf{r},k)=\sum^{6}_{\alpha=1}s_\alpha(\mathbf{r},k) \left(\mathbf{c}_\vartheta\cdot\mathbf{c}_\alpha\right)^m 
\end{equation}
where $\mathbf{s}(\mathbf{r},\!k)=[s_1(\mathbf{r},\!k), s_2(\mathbf{r},\!k), \cdots,s_6(\mathbf{r},\!k)]$ is the configuration of node $\mathbf{r}$ at time $k$, which is given by the number of particles $s_\alpha(\mathbf{r},k)$ occupying  channel $\alpha$ (with $\alpha \in [1, 2, \cdots, 6]$), 
 the strength of the velocity alignment is denoted by 
$\beta\in\mathbb{R}_+$, and $Z$ is the node normalization constant.  
Finally, $m=1$ corresponds to polar and $m=2$ to nematic velocity alignment. These transition probabilities are the volume exclusion-free analogues of those derived previously \cite{nava2017extracting}, which were obtained as discretized, steady state distributions of flocking off-lattice Langevin equation models with pairwise velocity alignment. Note that Eq.~\ref{eq:reorient} is a Maxwell-Boltzmann distribution, where the parameter $\beta$ is inversely proportional to the effective temperature. Indeed, as seen in \cite{nava2017extracting}, $\beta$ is proportional to the interaction strength, and inversely proportional to the noise amplitude of off-lattice, stochastic differential equation models. Thus, in the limit $\beta\rightarrow0$ (corresponding to infinitely high temperatures or noise amplitudes), Eq.~\ref{eq:reorient} reduces to a discrete, uniform distribution, with particles exhibiting no preference in their reorientation. On the other hand, in the limit $\beta\rightarrow\infty$ (corresponding to an absolute zero temperature or no noise), $T_{\vartheta}=1$ for a single orientation $\vartheta$ and $T_{\vartheta}=0$ for all other orientations, resulting in completely deterministic behavior, for $m=1$, and two opposite orientations with $T_{\vartheta}=\frac{1}{2}$, and $T_{\vartheta}=0$ for all other orientations, resulting in a one-dimensional, symmetric random walk along a single axis, for $m=2$.
We stress that Eqs.~(\ref{eq:reorient}) and~(\ref{eq:H}) define a purely local velocity alignment process: the reorientation dynamics does not consider the configuration of neighboring nodes.  
Note that the number of particles at node $\mathbf{r}$ at time $k$ is given by 
$n(\mathbf{r},k)=\sum_{\alpha}s_{\alpha}(\mathbf{r},\!k)$. 
This implies that Eq.~(\ref{eq:H}) for $m=1$ can be expressed as $H(\mathbf{r},k)_{\vartheta}=n(\mathbf{r},k) \mathbf{P}(\mathbf{r},k)\cdot \mathbf{c}_\vartheta$ 
where $\mathbf{P}(\mathbf{r},k)=\left[\sum_{\alpha}s_\alpha(\mathbf{r},k)  \mathbf{c}_{\alpha}\right]/n(\mathbf{r},k)$.

(ii) Particles move spatially according to the following rule: particles in node $\mathbf{r}$ at channel $\vartheta$ are transported into the same channel into the neighboring node $\mathbf{r}+\mathbf{c}_{\vartheta}$; see Fig.~\ref{fig:dynamics}.  It should be noted that lattice-gas cellular automata are mesoscopic, rather than microscopic, models. Therefore, time and space are coarse-grained. At this level of resolution, the reorientation processes that occur continuously in microscopic models are assumed to have reached a steady state at every discrete time step \cite{nava2017extracting}, and thus can only move at a fixed speed of one lattice site per time step.

\begin{figure}
    \centering
    \begin{subfigure}[b]{0.3\textwidth}
         \centering
         \includegraphics[width=\textwidth]{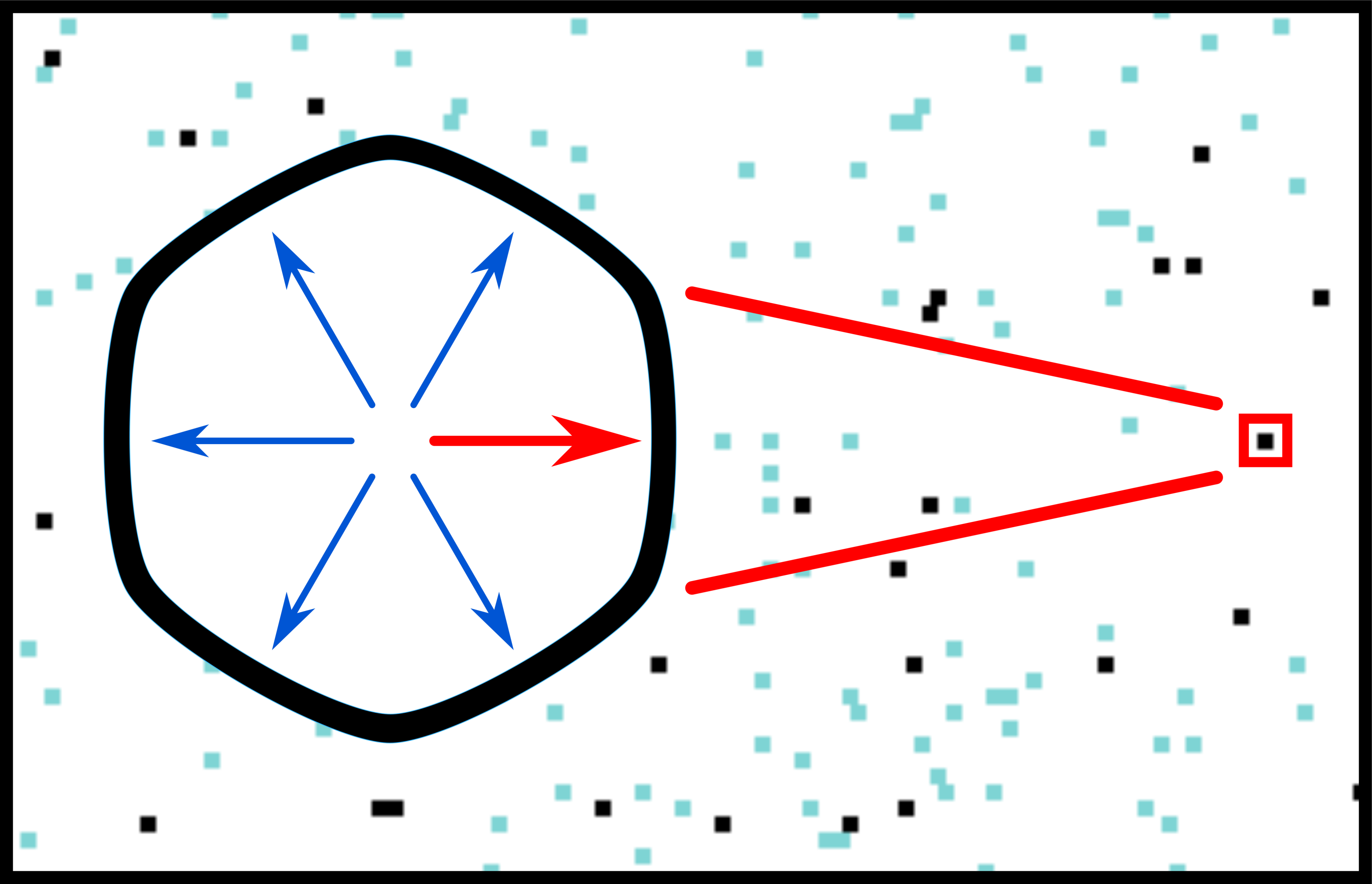}
         \caption{}
         \label{snapshotpol}
    \end{subfigure}
    ~
    \begin{subfigure}[b]{0.3\textwidth}
         \centering
         \includegraphics[width=\textwidth]{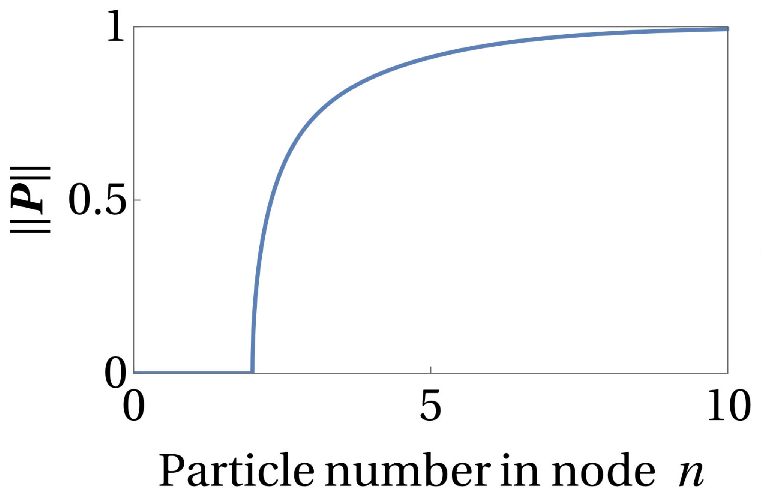}
         \caption{}
         \label{poltransition}
    \end{subfigure}
    ~
    \begin{subfigure}[b]{0.3\textwidth}
         \centering
         \includegraphics[width=\textwidth]{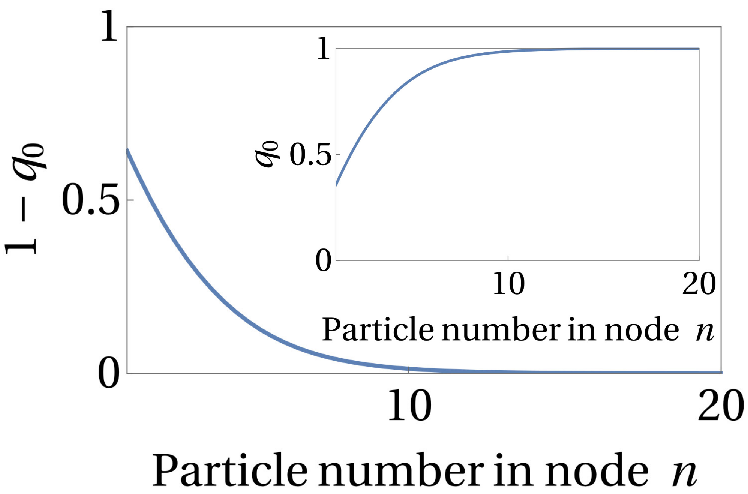}
         \caption{}
         \label{pstayleave}
    \end{subfigure}

    \begin{subfigure}[b]{0.3\textwidth}
         \centering
         \includegraphics[width=\textwidth, trim={0cm 1cm 0.5cm 0.5cm}, clip]{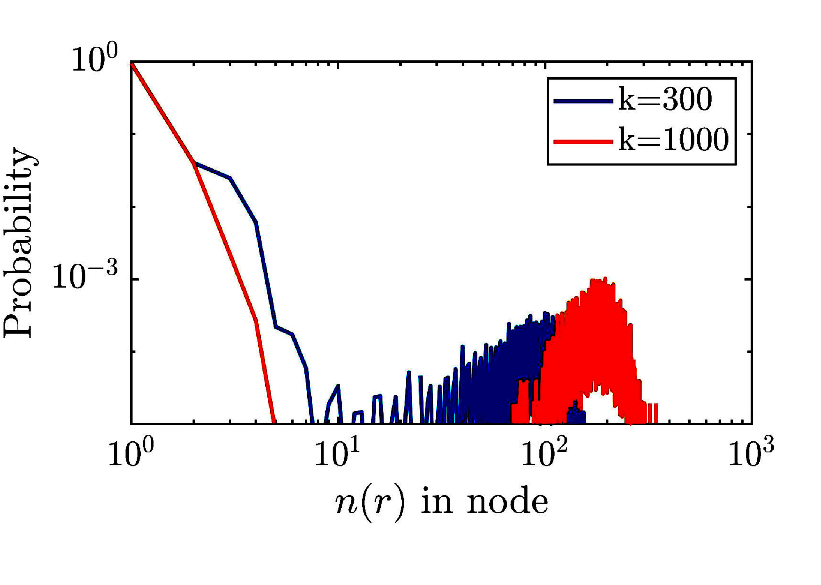}
         \caption{}
         \label{polcondens}
    \end{subfigure}
    ~
    \begin{subfigure}[b]{0.3\textwidth}
         \centering
         \includegraphics[width=\textwidth, trim={0cm 1cm 0cm 1cm}, clip]{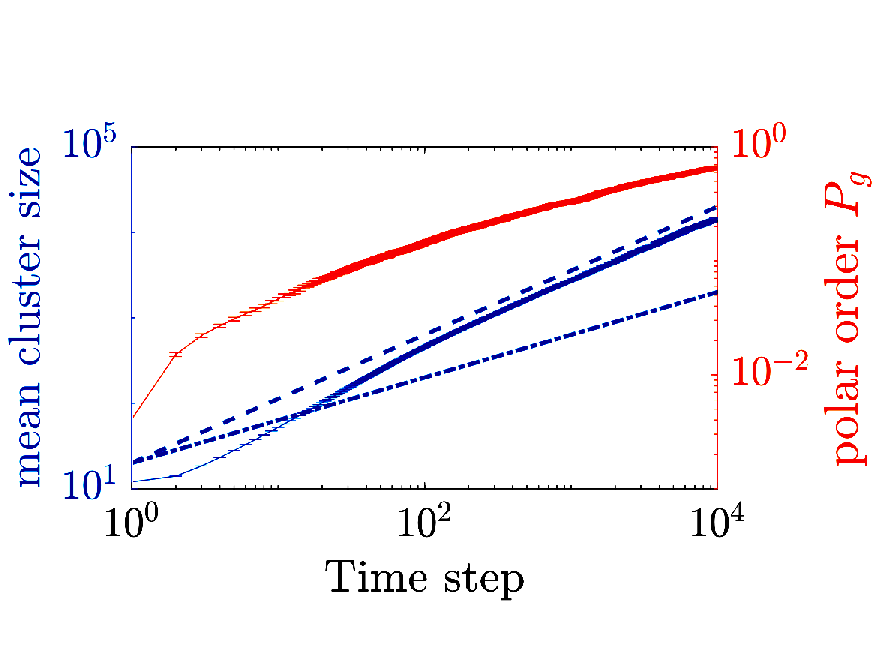}
         \caption{}
         \label{clustsize}
    \end{subfigure}
    ~
    \begin{subfigure}[b]{0.3\textwidth}
         \centering
         \includegraphics[width=\textwidth, trim={0.5cm 0.5cm 0.5cm 0cm}, clip]{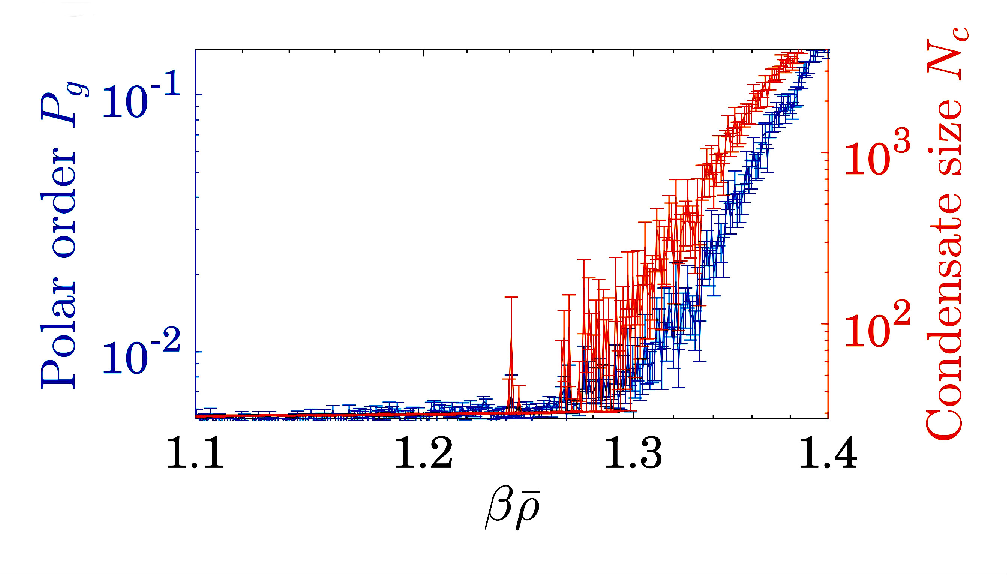}
         \caption{}
         \label{polorderncsize}
    \end{subfigure}

    \begin{subfigure}[b]{0.3\textwidth}
         \centering
         \includegraphics[width=\textwidth, trim={4.5cm 8.5cm 4cm 9cm}, clip]{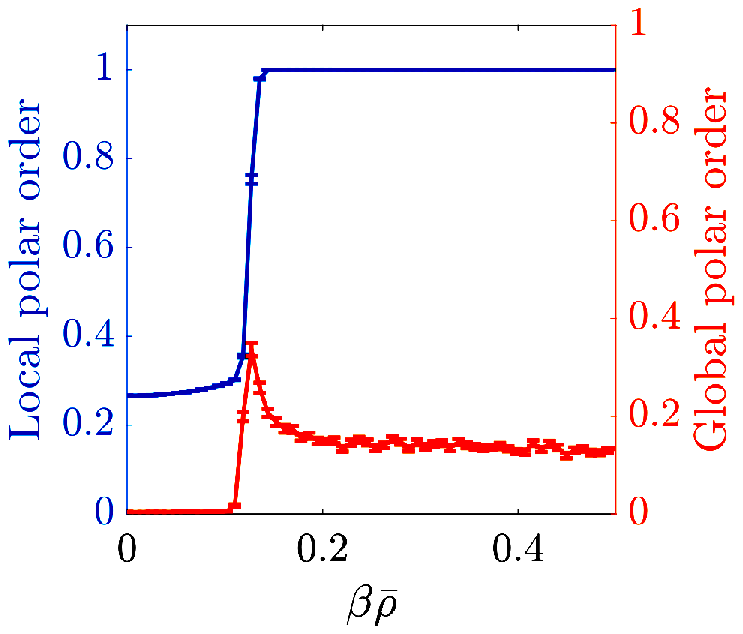}
         \caption{}
         \label{locvsglobpol}
    \end{subfigure}
    \caption{Dynamics of the polar, BA-LGCA. \protect\subref{snapshotpol}. Simulation snapshot; node occupancy is color coded, with dark color indicating large particle number. The inset shows the state of a node. See~\cite{SI} for movies. \protect\subref{poltransition}. Global polar order  $\left\|\mathbf{P}\right\|$ vs.  average occupation per node $n$.  \protect\subref{pstayleave}. Probability of a particle escaping a polarized condensate given its particle number. Inset:  Probability of a particle staying in a polarized cluster. \protect\subref{polcondens} Histogram of the number of particles in all lattice sites at two different time steps. \protect\subref{clustsize}. Mean cluster size (blue) and polar order  (red) as function of time; the dotted line corresponds to $t^{1/2}$, the characteristic growth in classic ZRPs \cite{evans2005nonequilibrium}, while the dashed line corresponds to $t^{3/4}$. Average over 200 simulations with $\beta=4.3$ and $\bar{\rho}=12$. \protect\subref{polorderncsize}. Polar order  $P_g$ and condensate size $N_c$ as a function of $\beta\bar{\rho}$ for average  $\bar{\rho}=12$. \protect\subref{locvsglobpol}. Global polar order $P_g$ and local polar order $P_{\ell}$ as a function of $\beta\bar{\rho}$ for average  $\bar{\rho}=12$. 150 simulations were averaged. A lattice size  of $60\times60$ was used for all simulations. Parameter values were recorded after 1000 time steps. }
    \label{fig:cond}
\end{figure}

\section{Emergence of polar condensates} 
For polar alignment ($m=1$), we observe, in strong contrast to other polar active systems
whose polar ordered phase consists of either a traveling polar band or an spatially homogeneous polar phase~\cite{vicsek_novel_1995, toner_long_1995, gregoire2004onset}, 
the emergence of polar, moving condensates; see  Fig.~\ref{snapshotpol}. 
In the following, we explain the mechanism that leads to the emergence of such condensates. 

We make use of a mean-field argument and define  the probability of finding a particle in channel $\vartheta$ of a certain node at time $k$ as $q_{\vartheta}(k) = s_{\vartheta}(k)/n(k)$, with $n(k)=\sum_{\alpha\in\Theta}s_{\alpha}(k)$ being the number of particles in the node.  
The node configuration, after reorientation, takes the form: 
\begin{eqnarray} \label{eq:q_eq}
q_{\vartheta}(k+1) = \frac{1}{Z} \exp\left[\beta n \sum_{\alpha \in  \Theta} \cos(\alpha - \vartheta) q_{\alpha}(k)\right]. 
\end{eqnarray}
Let us assume a steady state condition: $q_{\vartheta}(k+1)=q_{\vartheta}(k)$ for all $\vartheta$. 
Under this assumption, the transcendental Eq.~(\ref{eq:q_eq}) can be easily solved; see Figs.~\ref{poltransition} and \protect\subref{pstayleave}. 
Note that it is possible to recast Eq.(\ref{eq:q_eq}) as 
$q_{\vartheta}= \exp\left[\beta n \mathbf{P} \cdot \mathbf{c}_{\vartheta} \right]/Z$ in order to obtain an equation for the polar order parameter of the node: $\mathbf{P}  = \sum_{\alpha \in  \Theta} q_{\alpha} \mathbf{c}_{\alpha}$. 
And thus, $\mathbf{P}  = \sum_{\alpha \in  \Theta} \frac{1}{Z}\exp\left[\beta n \mathbf{P} \cdot \mathbf{c}_{\vartheta} \right]\mathbf{c}_{\vartheta}$. 
This transcendental equation  
describes a phase transition.   
The disorder state, given by  $||\mathbf{P}||=0$, and corresponding to 
$q_{\vartheta}=1/6$ for all $\vartheta$, is a solution of the above transcendental equation.  
This solution is stable for a number of particle $n$ below the critical $n^{*}$ value. 
This is evidenced by the behavior of  $||\mathbf{P}||$ in Fig.\ref{poltransition}. 
Above $n^{*}$, $q_{\vartheta}=1/6$ is not stable and particles 
accumulate preferentially in one velocity channel  and polar order in the node emerges. 
It is very important to realize that the probability of finding all particles in a node  occupying the same velocity channel, for $n>n^{*}$  
grows exponentially with the number of particles $n$ in the node. 
This implies that once all particles in a node occupy the same velocity channel, the probability that single particles transition to other velocity 
channels decays exponentially with $n$, as does  the probability of not finding the $n$ particles in the same node. 
This mechanism leads to a condensation of particles. 
To understand this, assume, without loss of generality, that all particles are initially located 
in channel $0$; then $q_0 = e^{\beta n}/Z = (1 + e^{\beta n (\gamma_1-1)} + \dots + e^{\beta n (\gamma_5-1)})^{-1}$, with $\gamma_j = \cos(-{\pi\,j/3})$ 
so $\gamma_j -1<0$.  
The probability that particles change channel is $1-q_0 = \sum_{j=1}^{5} e^{\beta n (\gamma_j -1)}/[1+\sum_{j=1}^{5} e^{\beta n (\gamma_j -1)}]$, 
which decays exponentially fast with $n$, 
while the probability of finding all particles again in channel $0$ after one time step behaves as $(1 + e^{\beta n (\gamma_1-1)} + \dots + e^{\beta n (\gamma_5-1)})^{-n}$. 
Consequently, the probability for a particle in a condensate of mass $n \gg n^{*}$  
to leave the condensate decays exponentially with $n$; Fig.~\ref{pstayleave}. 
This feature is shared by the standard zero-range process (ZRP)~\cite{evans2005nonequilibrium}.  
It should be mentioned that, in this case, randomness decreases exponentially with particle number $n$ due to the pairwise, additive definition of $H_{\vartheta}(\mathbf{r},k)$ (Eq.~\ref{eq:H}), which increases linearly as more aligned particles are present. Conversely, other population models may exhibit similar fluctuations originating from multiplicative noise terms~\cite{scheffer2017inequality}, which we do not consider in this model.
Particle condensation can be evidenced by looking at the distribution of  particles per node $P(n,t)$; Fig.~\ref{polcondens} (cf.~\cite{majumdar1998nonequilibrium}). 
The second peak that shifts over time is indicative of an aggregation dynamics. 
Note that here, it is the alignment mechanism that holds particles together. The emergent condensate moves ballistically in contrast to the processes studied in ~\cite{evans2005nonequilibrium, majumdar1998nonequilibrium}. 
This means that the condensation dynamics is not solely mediated by particle exchange with the bulk,  
but results from the  motion of the condensate:   as the condensate moves cross the system collects   clusters in its way and grows in size, Figs.~\ref{polcondens} and~\protect\subref{clustsize}. This is further supported by the fact that the mean condensation time increases with system size (see Fig.~S8 of the Supplementary Material), since clusters need to travel longer distances to accumulate particles in distant lattice sites.   
Note that, as argued in~\cite{peruani_kinetic_2013}, in the absence of velocity correlations among clusters (and only in this scenario), polar order becomes intimately connected with the number and distribution of moving clusters. 
As expected, in Fig.~\ref{polorderncsize}  we observe that as particles coalesce and the mean cluster size grows with time,  the global order parameter $P_g$ -- defined as $P_g  = || \sum_{\mathbf{r}}\sum_{\alpha \in  \Theta} q_{\alpha}(\mathbf{r}) \mathbf{c}_{\alpha} ||$ also does it. Furthermore, note 
that the mean cluster size scales with time with an exponent larger than $1/2$; cf. ~\cite{evans2005nonequilibrium}. 
Moreover, Fig.~\ref{polorderncsize} shows that the condensate size $N_c$, defined as $N_c=\left\langle\mathrm{max}\left\{s_\vartheta\left(\mathbf{r}\right):\vartheta\in\Theta,\mathbf{r}\in\mathcal{L}\right\}\right\rangle$, and $P_g$ exhibit almost identical dependency with $\beta \bar{\rho}$, with $\bar{\rho} =\left\langle n\left(\mathbf{r}\right)\right\rangle= N/L^2$, where $N$ is the number of particles in a system of linear size $L$. 
An additional observable obtained from simulations was the local polar order, defined as $P_{\ell}=\sum_{\mathbf{r}}||\sum_{\alpha\in\Theta}q_{\alpha}(\mathbf{r})\mathbf{c}_{\alpha}||/N_o$, where $N_o$ is the number of nodes $r\in\mathcal{L}$ with at least one particle in any channel. Thus, $P_{\ell}$ is the average of the polar order within individual nodes, across all occupied nodes in the lattice. It is evident that, if particles are homogeneously distributed among all channels in every non-empty node, then $P_{\ell}=0$. Conversely, if for every node, all particles are contained within a single channel, then $P_{\ell}=1$. Fig.~\ref{locvsglobpol} shows a plot of the global order $P_g$ and the local polar order $P_{\ell}$, as a function of the model parameters after several time steps. After the transition, the local polar order $P_{\ell}$ saturates up to its maximum value, while the global polar order $P_g$ exhibits non-monotonic behavior, reaching a maximum exactly at the transition point. This means that, after the transition, most occupied nodes accumulate particles within a single node. At the transition point, few condensates form and therefore the condensate with the most particles increases the overall global polar order. After the transition, several condensates with their own orientation form at once. The difference among the condensates' orientation drives the goal polar order down. Thus, disconnected condensates are present for all parameter values after the transition.

\section{Emergence of nematic filaments} 
\begin{figure}
    \centering
    \begin{subfigure}[b]{0.4\textwidth}
         \centering
         \includegraphics[width=\textwidth]{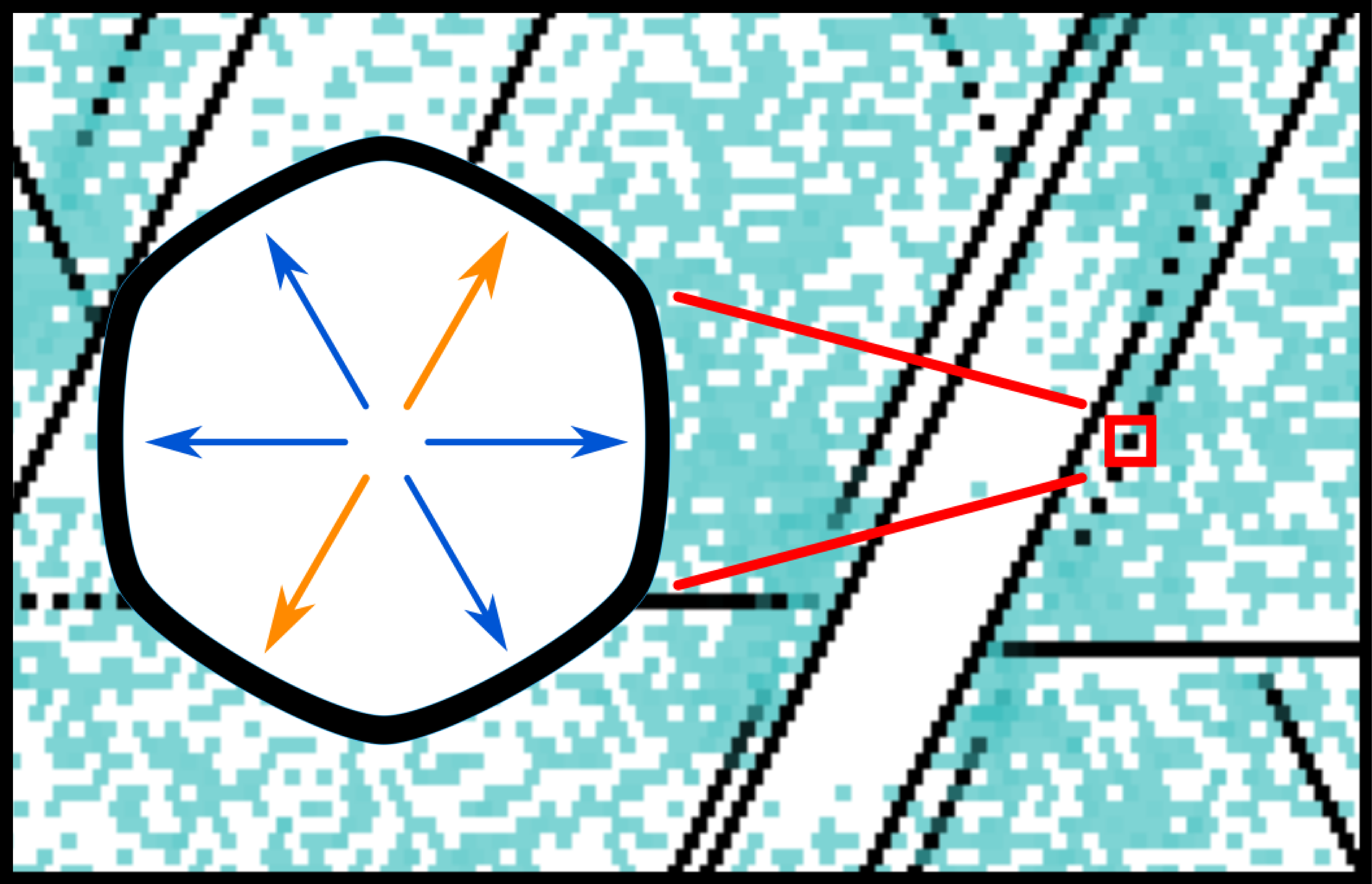}
         \caption{}
         \label{snapshotnem}
    \end{subfigure}
    ~
    \begin{subfigure}[b]{0.4\textwidth}
         \centering
         \includegraphics[width=\textwidth, trim={4.5cm 8.5cm 4cm 9cm}, clip]{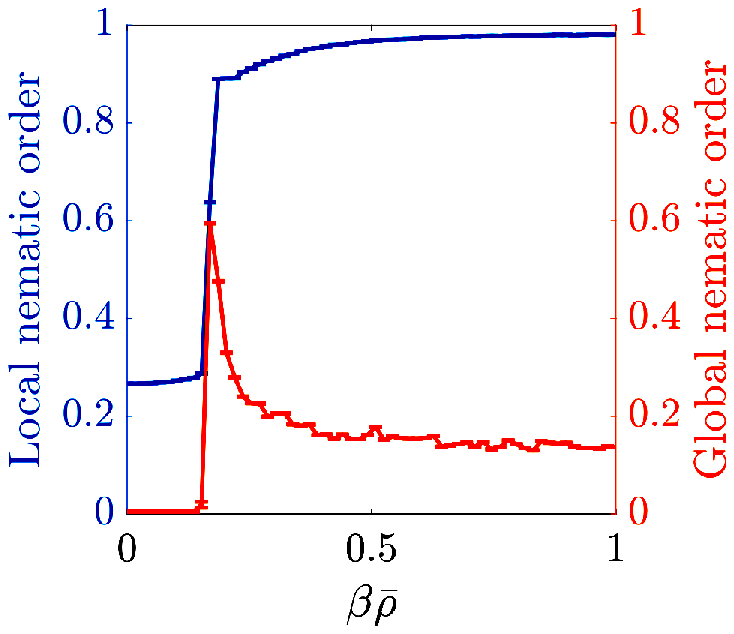}
         \caption{}
         \label{globallocalnematic}
    \end{subfigure}
    
    \begin{subfigure}[b]{0.4\textwidth}
         \centering
         \includegraphics[width=\textwidth]{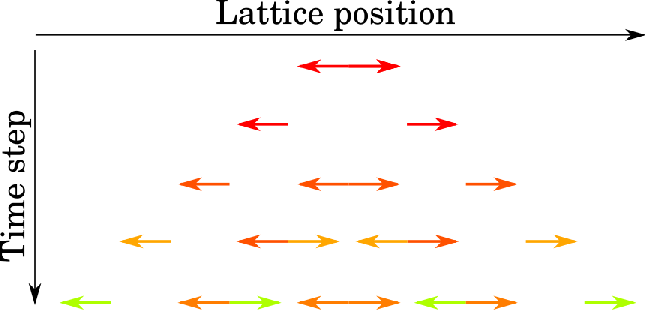}
         \caption{}
         \label{bandform}
    \end{subfigure}
    ~
    \begin{subfigure}[b]{0.4\textwidth}
         \centering
         \includegraphics[width=\textwidth]{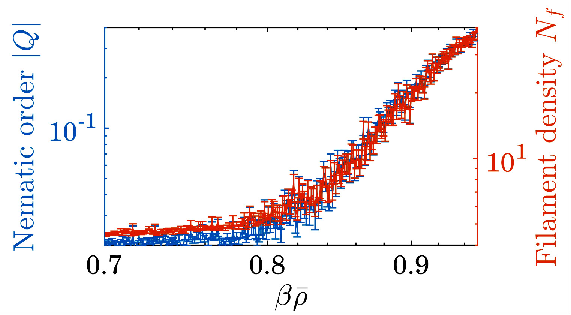}
         \caption{}
         \label{dennemord}
    \end{subfigure}
\caption{Dynamics of the nematic, BA-LGCA. \protect\subref{snapshotnem}. Snapshot of a simulation, with high node occupation numbers displayed in dark color; inset shows the state of a node. See~\cite{SI} for movies. \protect\subref{globallocalnematic}. Global nematic order $|Q|$ and local nematic order $Q_{\ell}$ as a function of $\beta\bar{\rho}$ for average  $\bar{\rho}=12$. \protect\subref{bandform}. Schematic depicts the filament formation process. Arrows indicate the direction of particle groups, with colors, from red to magenta, decreasing particle number. \protect\subref{dennemord}. Filament density vs. $\beta\bar{\rho}$ averaged over 150 simulations for fixed $\bar{\rho}=0.6$. a Lattice of size $60\times60$ was used for all simulations. Parameter values were recorded after 1000 time steps.}
\label{fig:band}
\end{figure}

The strict local nature imposed by zero-range interactions seems to prevent nodes to exchange  orientational information in order to establish extended ordered structures. 
This is true for polar alignment: the polar order of two neighboring nodes is not coupled.  
Counterintuitively, for nematic alignment, and in contrast to polar alignment, nodes can exchange orientational information and  particles self-organize 
into stable, elongated, nematic filaments that lead to global nematic order; Fig.~\ref{snapshotnem}. 
We stress that these nematic filaments should not be confused with, and are different from the nematic bands 
observed in off-lattice models~\cite{ginelli_large_2010}: 
i) the width of these filaments is always of one node and does not vary with the distance to the critical point as occurs in nematic bands~\cite{ginelli_large_2010}, and ii) as consequence of this, the transversal instability observed for nematic bands, precluding global nematic order, is not present~\cite{grossmann_mesoscale_2016, peshkov_nonlinear_2012}.  

How can an extended nematic filament emerge from zero-range interactions? 
Picture an initial configuration where a large number $n$ of particles are placed on a single channel of a lattice site. As in the polar case, $q_{\vartheta}=1/6$ is not stable for $n>n^{*}$ and particles 
  accumulate  in two opposite velocity channels. 
In the next time step, particles  move to neighboring sites associated to these two velocity channels.
If the density in each new lattice site is high, the same process occurs in the new sites,  and thus information on the initial orientational order, propagates forward and backward, as shown in the sketch of Fig.~\ref{bandform}. 
If we consider that this process starts taking place in an area surrounded by a disordered population  of particles, then as the nematic filament expands enlarging its scattering cross-section,  these particles  feed 
the growing  filament until it eventually percolates. 
The described mechanism leads to  global nematic order as evidenced by the global nematic order parameter  $|Q|=\left|\sum_{\mathbf{r}}\sum_{\alpha\in\Theta}q_{\alpha}\left(\mathbf{r}\right)e^{2i\alpha}\right|$.  
Note that the emergent filament contains a macroscopic fraction of the particles in the system, as  evidenced by the behavior of the filament density $N_f$ with $\beta$, which is similar to the one of $N_c$ for polar alignment;  Fig.~\ref{dennemord}.  
Furthermore, we observe that for fixed $\bar{\rho}$,  the $N$ grows as $L^2$, and thus $N_f$ as  $L$, which implies that for $L \to \infty$, this quantity diverges: the nematic filament acts as a condensate with a macroscopic particle fraction located at it moving in two opposite directions. 
The local nematic order, $Q_{\ell}=\sum_{\mathbf{r}}\left|\sum_{\alpha\in\Theta}q_{\alpha}\left(\mathbf{r}\right)e^{2i\alpha}\right|/N_o$ was also observed in simulations. Here as well, we observed a monotonic increase of the local nematic order and a non-monotonic behavior of the global nematic order; Fig.~\ref{globallocalnematic}. As before, this indicates the formation of few filaments with most particles at the transition point, driving polar order up, and several filaments with different orientations far above the transition point, mantaining a high local nematic order, but inhibiting nematic order at a global level.

\section{Hydrodynamic equations} 
The onset of the (macroscopic) symmetry breaking, while the system remains near  the disordered phase, can be described in an approximate manner by calculating the expected occupation numbers of lattice velocity channels. However, note that well-developed condensates and filaments cannot be described by smooth PDE solutions.  
As a first step, let us define $f_\vartheta\left(\mathbf{r},k\right)\coloneqq\left\langle s_\vartheta\left(\mathbf{r},k\right)\right\rangle$. 
By performing an in-node mean-field approximation and employing the binomial statistics of velocity channel occupation numbers, one 
obtains the lattice-Boltzmann equation (LBE):  $f_{\vartheta}\left(\mathbf{r}+\mathbf{c}_{\vartheta},k+1\right)=\rho\left(\mathbf{r},k\right)T_{\vartheta}\left(\mathbf{r},k\right),$ where $\rho\left(\mathbf{r},k\right)\coloneqq\sum_{\vartheta\in\Theta}f_\vartheta\left(\mathbf{r},k\right)=\left\langle n(\mathbf{r},k)\right\rangle$ is the density, equal to the expected value of the total number of particles in each lattice site,  and $T_{\vartheta}\left(\mathbf{r},k\right)\coloneqq\frac{1}{Z}\exp\left[\beta\sum_{\alpha\in\Theta}f_\alpha\left(\mathbf{r},k\right)\left(\mathbf{c}_\vartheta\cdot\mathbf{c}_\alpha\right)^m\right]$ is the mean-field alignment probability. 
We perform a discrete Fourier transform on $\vartheta$ to arrive to hydrodynamic fields 
that depend on position and time. 
For polar alignment ($m=1$), the relevant hydrodynamic fields reduce to 
 the local density $\rho\coloneqq\sum_{\vartheta\in\Theta}f_\vartheta$ and local polar order (or momentum) $\mathbf{M}\coloneqq \sum_{\vartheta\in\Theta} f_\vartheta \left(\cos\vartheta,\sin\vartheta\right)$. 
The temporal evolution of these fields, in dimensionless form, leading to unit active speed, is given by:
\begin{subequations}
\label{eq:polar}
\begin{align}
& \partial_t\rho+\nabla\cdot\mathbf{M}=0 \\
\label{eqM}
\begin{split}
\partial_t\mathbf{M}+\frac{1}{2}\left[\nabla\varrho-\mathcal{D}\left(\nabla\varrho\right)\right]+ \\ \frac{\beta}{\rho}\left[\left(\mathbf{M}\cdot\nabla\right)\mathbf{M}-\frac{1}{2}\nabla M^2+\left(\nabla\cdot\mathbf{M}\right)\mathbf{M}\right] \\= \rho\left\langle\mathbf{u}\right\rangle-\mathbf{M},
\end{split}
\end{align}
\end{subequations}
where $\langle \mathbf{u} \rangle\coloneqq \sum_{\vartheta\in\Theta} T_\vartheta \left(\cos\vartheta,\sin\vartheta\right)$, $\mathcal{D}\left(\nabla\rho\right)\coloneqq\frac{1}{\rho^2}\left[\mathbf{M}\left(\mathbf{M}\cdot\nabla\rho\right)+\mathbf{M}\times\left(\mathbf{M}\times\nabla\rho\right)\right]$. 
For detailed derivation, see~\cite{SI}. 
The term $\langle \mathbf{u} \rangle$ can be approximated, assuming $\beta \ll 1$, by  
a Taylor series. 
If we approximate it up to third order in $\beta$, we find 
$\left\langle\mathbf{u}\right\rangle=\frac{\beta}{2}\mathbf{M}\left(1-\frac{\beta^2}{8}\mathbf{M}^2\right)$, and by replacing this expression into Eq.~(\ref{eqM}), we obtain an equation that is invariant under arbitrary rotations of $\mathbf{M}$ as occurs in Toner-Tu theory~\cite{toner_long_1995}. 
However, if the expansions is performed up to 5th order, we uncover 
that steady solution of the form 
$\mathbf{M}=M (\cos \theta, \sin \theta)$, with $M$ a scalar constant and $\theta$ an angle, cannot exist for arbitrary values of $\theta$. 
The only possible solutions correspond to $\theta = \frac{a \pi}{3}$, with $a\in [0, 1, \cdots,5]$, i.e. to one of the possible 6 directions of the lattice; for details see~\cite{SI}.   

Differences with the standard off-lattice Vicsek model can be shown even using the third order expansion in $\beta$ mentioned above. There are, as expected, two spatially homogeneous solutions, corresponding to $M=0$ (disorder) and $M>0$ (order).  Performing a linear stability analysis around the disordered homogeneous steady state, and assuming perturbations along a single direction, we find that  the system is stable against uniform perturbations for an initial average density $\beta \bar{\rho}<2$, showing that the product $\beta\bar{\rho}$ is the effective bifurcation parameter. 
The second steady state corresponds to spatially homogeneous polar order state with  
 $M=\left[8\left(1-\frac{2}{\bar{\beta\rho}}\right)\right]^{1/2}$. The ordered homogeneous state only exists when $\beta \bar{\rho}>2$, where the disordered state becomes unstable.  
For  non-uniform perturbations,  the spatially homogeneous ordered state, that is never observed in the LGCA,  is stable only for the parameter range $2.564\lessapprox\beta\bar{\rho}<3$. Otherwise,  the real parts of the eigenvalues grow with increasing wavenumber, indicating that small wavelength perturbations dominate, as confirmed by direct integration of the hydrodynamic equations. This instability indicates a drastic density increase in localized areas of the space, which can be associated to the emergence of condensates at this level of description. Numerical evidence strongly suggests that spatially homogeneous ordered states, and ordered polar bands are not possible in the LGCA model, even when when forcing the system by using these configurations as initial conditions (see SI).

For nematic alignment ($m=2$), the relevant hydrodynamic fields  
include nematic order, defined for simplicity as 
$Q \coloneqq\sum_{\vartheta\in\Theta}f_\vartheta \left( \cos2\vartheta+i\sin2\vartheta\right)$, 
and the hydrodynamic equations read: 
\begin{subequations}
\label{eq:nema}
\begin{align}
& \partial_t\rho=\mathfrak{d}\mathfrak{d}^{\ast}\rho+\mathfrak{R}\left(\mathfrak{d}^2Q^{\ast}\right) \\
\begin{split}
\partial_tQ=\frac{1}{2}\left\{\mathfrak{d}^2\rho+\mathfrak{d}\mathfrak{d}^{\ast}Q+2\mathfrak{d}^{\ast}\left[\mathfrak{R}\left(\mathfrak{d}Q\right)\right]\right\} \\+\frac{\beta \rho}{4}\left[Q+\frac{\beta}{8}Q^{\ast}\left(Q^{\ast}-\frac{\beta}{4}Q^2\right)\right]-Q,
\end{split}
\end{align}
\end{subequations}
where $\mathfrak{d}\coloneqq\partial_x+i\partial_y$, ${}^{\ast}$ indicates  complex conjugate, $\mathfrak{R}\left(\cdot\right)$ denotes real part, and fast relaxation of $\mathbf{M}$ was assumed.  
It is worth stressing, as we explain below, that according to Eq.~(\ref{eq:nema}) nematic order is constrained to exist, as occurs for $\mathbf{M}$, on the orientations allowed by the lattice structure. 
Eq.~(\ref{eq:nema}) possesses spatially homogeneous disordered and nematically ordered solutions. 
The disordered state is stable when $\beta \bar{\rho}<4$, where the bifurcation depends only on the product $\beta\bar{\rho}$, as before. 
The order in the spatially homogeneous steady state is given by 
 $Q=\left(2\pm2\sqrt{9-\frac{32}{\beta \bar{\rho}}}\right)e^{i\frac{2k\pi}{3}}$, $k\in\left\{0,1,2\right\}$. 
 Two features are worth discussing.  First, the ordered state exists when $\beta \bar{\rho}>\frac{32}{9}$, i.e. it exists even when the disordered state is stable, indicating a subcritical bifurcation. Second, while for $\frac{32}{9}<\beta \bar{\rho}<4$ three pairs of solutions with different moduli, but same argument $\mathrm{arg}\left(Q\right)\in\left\{\frac{2k\pi}{3},k\in\mathbb{N}\right\}$ exist, for $\beta \bar{\rho}>4$, six different solutions exist, each one with a different argument corresponding to one of the lattice directions $\vartheta\in\Theta$. 
For $\beta \bar{\rho}>4$, solutions with $\mathrm{arg}\left(Q\right)=\frac{2k\pi}{3}$ are stable, while those with $\mathrm{arg}\left(Q\right)=\frac{\left(2k+1\right)\pi}{3}$ are unstable. This is a consequence of the constraints imposed on $Q$ by the lattice, such that nematic structures can only emerge along the three hexagonal lattice axes, as observed in the LGCA simulations. Again, simulations suggest that spatially homogeneous ordered states, as well as spatially extended bands are not stable, and consistently break into nematic filaments, or revert back to the homogeneous disordered state. In this case, however, ordered states are much less unstable than in the polar case, and need long times in order to break apart (see SI).

\section{Discussion} 
The original  LGCA was introduced with the intention of reproducing basic 
fluid mechanics properties. 
However, the initial study performed on a square lattice~\cite{hardy1976molecular} displayed  
a series of anomalies due to the use a finite number of velocity channels. 
Later on, it was found that 
such anomalies vanish in hexagonal lattices~\cite{frisch1986lattice}, 
where  the phenomenology of Navier–Stokes equations is recovered, despite the use of only 6 velocity channels. 
Similarly, it is tempting to speculate that an active LGCA with polar velocity alignment in an hexagonal lattice (and short-range interactions) has to reproduce the large-scale properties  
predicted by the Toner-Tu equations~\cite{toner_long_1995} for a (compressible) active polar fluid. 
After all, it is expected that most microscopic details are, at macroscopic scales, irrelevant except for the range and symmetry of the interactions.  
However, our study shows that spatially homogeneous ordered phases (cf. Toner-Tu phase~\cite{toner_long_1995}) 
or polar bands (cf.~~\cite{gregoire2004onset, solon2015flocking}) cannot exist
in the LGCA model defined in this work.  
Moreover, in this LGCA, order emerges via a fundamentally different phase transition from the ones previously reported~\cite{vicsek_novel_1995, toner_long_1995, gregoire2004onset,nesbitt2021uncovering, solon2015flocking, bussemaker1997mean, ginelli_large_2010} in active matter. 
%|
In LGCA, the emergence of global order is intrinsically related to a condensation transition -- that occurs in the absence of  an external potential -- which shares similarities with the ZRP. 
For the polar LGCA, it is associated to the formation of a polar macroscopic condensate, 
while for the nematic LGCA to the collapse of a macrocopic number of particles in the systems into 
a nematic filament (that leads to global nematic order; cf.~\cite{ginelli_large_2010, peshkov_nonlinear_2012}).  

After their introduction as fluid-dynamical models, LGCA have been used to model biological systems, specifically for modeling collective cell migration (see~\cite{deutsch2021bio} for a review). The original LGCA formulation considers an exclusion principle, which limits the maximum number of cells which may be present at a single spatial point at a given time. On the one hand, our model could be considered as a 2D projection of a 3D system. For example, the accumulation of particles at certain lattice sites could correspond to the formation of an elongated structure by the piling up of agents, such as during the fruiting body and slug formation in \emph{D.~ discoideum}~\cite{maree2001amoeboids}. On the other hand, cells are quite compressible, and are able to accumulate in great numbers. Experimentally, it has been found that cells can greatly reduce their volume during tissue growth~\cite{puliafito2012collective,aland2015mechanistic}. Cells also regularly change their size to maintain epithelial homeostasis~\cite{gradeci2021cell}. For example, the photoreceptor epithelium shows a wide variety of receptor packing densities, corresponding to photoreceptor cells of varying sizes~\cite{jiao2014avian}. Furthermore, invasive cancer cells can be ten times as compressible as their healthy counterparts~\cite{swaminathan2011mechanical}, and cell compressibility has been shown to correlate with invasive potential~\cite{cheburkanov2022imaging}. Moreover, it has been shown that densely packed cells confined to narrow environments can still move collectively, as long as they mantain sufficiently high adhesive contacts and velocity correlation levels~\cite{ilina2020cell}. The present LGCA model extension, which does not consider an exclusion principle, represents a first step towards a better mathematical representation of these cellular aggregation and compression processes.

Finally, we strongly suspect that the condensation reported here is  related 
to the recent findings that indicate that here exist 
 fundamental differences between additive and non-additive interactions in vectorial active matter~\cite{chepizhko2021revisiting, kursten2021quantitative}. 
In particular, we speculate that the emergent macroscopic behavior can be affected, 
not only by modifying the symmetry or range of the interactions (see~\cite{bussemaker1997mean} for an analogous model with extended range and an exclusion principle), as it is usually expected, but 
by eliminating the dependency on $n(\mathbf{r},k)$ in Eq.~(\ref{eq:H}), 
for instance by defining for $m=1$, $H(\mathbf{r},k)_{\vartheta}= \mathbf{P}(\mathbf{r},k)\cdot \mathbf{c}_\vartheta$. 
This will be the focus of future investigations.

\section*{Author Contributions}
JMNS: Formal analysis, investigation, methodology, software, visualization, writing (original draft), and writing (review and editing). HH: Funding acquisition, supervision, and writing (review and editing). AVB and LB: supervision, methodology, and writing (review and editing). AD: Funding acquisition, supervision, project administration, and writing (review and editing). FP: Conceptualization, formal analysis, investigation, methodology, project administration, supervision, and writing (review and editing).

\section*{Conflicts of interest}
There are no conflicts to declare.

\section*{Acknowledgements}
F.P. acknowledges financial support from CY Initiative through the Ambition project ``Collective Intelligence" and 
Labex MME-DII.
JMNS acknowledges support from the PAPIIT-UNAM grant, project IA104821.

% \printbibliography

\end{document}